# Quality of Open Source Systems from Product Metrics Perspective

Mamdouh Alenezi[1] and Ibrahim Abunadi[2]

[1] [2] **College of Computer & Information Sciences**
**Prince Sultan University, Riyadh, Saudi Arabia.**

**Abstract**

Software engineering and information systems practices seek ultimately to create the flawless product. One of the tools used to improve the quality of software development is the use of metrics. In this paper, metrics retrieved from open source software were analyzed for quality attributes. Defect density is considered a strong indication of the quality of software product. Few studies have taken into consideration the density of defects while looking into quality of software and proneness to defects. Analysis of this study has shown that defect density is relevant to different developers and different product sizes. Thus, open source project has shown to have low defect density and the larger the product the lower the defect density is. In addition, this study has shown that there are different metrics that correlate with each other indicating that some of these metrics have conceptual and practical relevance to each other. Another relationship was tested between the number of bugs and the metrics. Results indicated that most attributes had positive correlation with the number of bugs with exception to coupling between cohesion among methods of class.

**Keywords:** *Software Quality, Software Metrics, Open Source, Defect Density*

## 1. Introduction

One of the essential objectives of the software engineering and information systems discipline is to develop techniques and tools for high-quality software solutions that are stable and maintainable. Software managers and developers use several measures to measure and improve the quality of a software solution throughout the development process. These measures assess the quality of different software attributes, such as product size, cohesion, coupling, and complexity. Researchers and practitioners use software metrics to understand and improve software solutions and the processes used to develop them. Determining the relationship between software metrics aids in clarifying practical issues with regard to the relationship between the quality of internal and external software attributes. Moreover, this understanding helps software practitioners and engineers to determine the factors that should be considered during the quality-assessment process.

The attributes of software quality can be categorized into two main types: internal and external. Internal quality attributes can be measured using only the knowledge of the software artifacts, such as the source code, whereas the measurement of external quality attributes requires the knowledge of other factors, such as testability and maintainability. The attributes of software quality, such as defect density and failure rate, are external measures of the software product and its development process. The focus of this paper is on internal attributes.

The field of software metrics has two main requirements: 1) enabling software developers to manage the software development process. For example, developers need to determine the resources or time needed to deliver; and 2) enabling researchers to define and measure software attributes objectively in order to gain a better understanding of software engineering [1]. The concepts of software metrics are coherent, understandable, and well established. Therefore, it is useful to develop and evaluate the quality of software solutions using these metrics. Metrics are measures of different aspects of an endeavor, and they help software engineers to determine whether they are progressing toward the goal of that endeavor.

Software metrics are used to measure the degree to which a software system possesses a certain property. There are three categories of software metrics. This classification is based on what they measure and the area of software development on which they focus. At a very high level, software metrics can be classified as process metrics, project metrics, and product metrics [2]: 1) process metrics are used to improve software development and maintenance; 3) project metrics describe the project's characteristics and execution, such as explaining the cost, schedule, productivity the number of software developers, and the staffing pattern over the life cycle of the software; 3) product metrics describe the characteristics of the product, such as size, complexity, design features, performance, and quality level.

One of the main goals of software engineering research is to provide evidence to support practitioners and facilitate them in making correct decisions during the development of the software [1]. Reaching these decisions always depends on how the data are analyzed and which information is extracted from the data during the analysis. In this paper, we examine the product metrics of several open source systems in order to determine the quality of these systems and how they compare to each other.

In this work, we empirically analyze the quality of several open source software systems. The remainder of this paper is organized as follows. Section 2 discusses quality definitions. Section 3 describes the data used in this study. Section 4 discusses the methodology, and Section 5 evaluates the experimental data. The empirical study is described in Section 5. Related work is discussed in Section 6. The conclusions are presented in Section 7.

## 2. Quality

Quality is defined variously depending on the context. We survey the definitions that are the best understood by the following international organizations:

- The German Industry Standards DIN 55350 Part 11 defines quality as "Quality comprises all characteristics and significant features of a product or an activity which relate to the satisfying of given requirements."
- The ANSI Standard ANSI/ASQC A3/1978 defines quality as "the totality of features and characteristics of a product or a service that bear on its ability to satisfy the given needs."
- The IEEE Standard (IEEE Std. 729-1983) defines quality as "The totality of features and characteristics of a software product that bear on its ability to satisfy given needs: for example, to conform to specifications; the degree to which software possesses a desired combination of attributes; the degree to which a customer or user perceives that software meets his or her composite expectations; the composite characteristics of software that determine the degree to which the software in use will meet the expectations of the customer."
- Pressman [3] defines the software quality in terms of the conformance to explicitly stated functional and performance requirements, explicitly documented development standards, and implicit characteristics that are expected of all professionally developed software
- The IEEE definition of Software Quality focuses on customer satisfaction, and the degree to which a system, component, or process meets specified requirements
- The IEEE definition of "Software Quality" focuses on the fulfillment of requirements, that is, the degree to which a system, component, or process meets the customer's or user's needs or expectations
- In addition to these definitions, software quality is usually dependent on the context in which it is required. Hence, in this work, we use the quality measure of defect density, which is usually defined as the number of defects found divided by size. One of the measures of the software size that is widely used in the open source community is the number of lines of codes in thousands, Kilo Lines of Codes, or KLOC, which is used in this paper.

## 3. Dataset

We conducted an empirical study on eight open source systems. We used several criteria to select the systems: 1) well-known systems that are used very widely; 2) sizable systems that yield realistic data; 2) actively maintained systems; 4) systems with publically available data, which is crucial in empirical studies. Table 1 shows the descriptive statistics of the dataset.

Table 1. Selected Software Systems

| System | Ver | Classes | KLOC | # of Bugs |
|---|---|---|---|---|
| Camel | 1.6 | 965 | 113 | 500 |
| Xalan | 2.7 | 909 | 428.5 | 1213 |
| Tomcat | 6.0.389418 | 858 | 300.6 | 114 |
| Ant | 1.7 | 745 | 208.6 | 338 |
| Xerces | 1.4.4 | 588 | 141.2 | 1596 |
| jEdit | 4.3 | 492 | 202.3 | 12 |
| POI | 3.0 | 442 | 129.3 | 500 |
| Velocity | 1.6.1 | 229 | 57 | 190 |

In this study, we used the dataset collected by [4], which is available online at the PROMISE repository. The systems in this dataset are as follows: Camel, Xalan, Tomcat, Ant, Xerces, jEdit, POI, and Velocity. Apache Camel is a powerful open source integration framework based on known Enterprise Integration Patterns with powerful Bean Integration. Xalan is a software library that implements the XSLT 1.0 XML transformation language and the XPath 1.0 language. The Xalan XSLT processor is available for both the Java and C++ programming languages. Tomcat is web server and servlet container. It implements several Java EE specifications, including Java Servlet, JavaServer Pages (JSP), Java EL, and WebSocket. Ant is a software tool used to automate software-building processes. It is similar to Make, but it is implemented using the Java language and requires the Java platform; it is best suited for building Java projects. Xerces is a parser that supports the XML 1.0 recommendation and contains advanced parser functionality, such as support for XML Schema 1.0, DOM level 2, and SAX version 2. jEdit is a mature programmer's text editor supported by hundreds (including the time-developing plugins) of person-years of development. It is written in Java and runs on any operating system that supports Java, including Windows, Linux, Mac OS X, and BSD. The POI project consists of APIs that are used to manipulate various file formats based on Microsoft's OLE 2 Compound Document format, and the Office OpenXML format, which uses pure Java. Velocity is a Java-based template engine that provides a template language that is used to reference objects defined in Java code. It aims to ensure the clean separation between the presentation tier and business tiers in a Web application.

The metrics are categorized as follows: coupling, cohesion, inheritance, and product size. The metrics were derived from several suites of metrics. We focus on object-oriented metrics because they are accessible in the early stages of software development. The selected metrics of open source software systems are shown in Table 2. These metrics have been widely studied in the literature [5, 6, 7, 8, 9].

Table 2. Metrics Names

| *Metric Name* |
| --- |
| Weighted methods per class (WMC) |
| Depth of Inheritance Tree (DIT) |
| Number of Children (NOC) |
| Coupling between object classes (CBO) |
| Response for a Class (RFC) |
| Lack of cohesion in methods (LCOM) |
| Lack of cohesion in methods (LCOM3) |
| Afferent couplings (Ca) |
| Efferent couplings (Ce) |
| Number of Public Methods (NPM) |
| Data Access Metric (DAM) |
| Measure of Aggregation (MOA) |
| Measure of Functional Abstraction (MFA) |
| Cohesion Among Methods of Class (CAM) |
| Inheritance Coupling (IC) |
| Coupling Between Methods (CBM) |
| Average Method Complexity (AMC) |
| McCabe's cyclomatic complexity (CC) |
| Lines of Code (LOC) |

## 4. Methodology

### 4.1 Correlations of the Metrics

To understand the relationships between software metrics, their correlation coefficients (i.e., the strength of relationships among their counterparts) are measured. We use the correlation between the metrics in order to find redundant metrics. Metrics that correlate measure similar aspects of software modules. We used Kendall's nonparametric measure of rank correlation [10]. Our choice is justified as follows: Pearson's correlation coefficients are highly influenced by outliers; and Spearman's rank correlation coefficient includes many equal values found in integer data [11].

### 4.2 Defect Density Evaluation

Defect density is one of the most established measures of software quality [12]. Defect density consists of post-release defects per thousand lines of a delivered code [13]. This definition is used mainly among practitioners to calculate and evaluate the quality of their projects at a certain phase of development. Defect density is used to measure the quality of the software product. It indicates the improvements in the quality of the successive releases of certain software. The lower the number of defect densities, the better the software quality is. Defect density can be computed using Eq 1 as follows:

$$\text{Defect Density} = \frac{\text{Number of Defects}}{\text{KLOC}} \qquad (1)$$

Defect density is jointly correlated with several developers and software sizes [14]. The size of the project is an influential factor (i.e., large projects have lower defect density). The mode of development mode is another factor that affects the defect density rate (i.e., open source projects have a lower defect density)[13].

## 5. Experimental Evaluation

### 5.1 Correlations of the Metrics

To study the relationships and correlations among the 19 metrics, we computed their cross-correlation values. The results are shown in Table 3 where the absolute values above 0.6 are highlighted in bold. We found a high correlation between several pairs of metrics. RFC was fairly correlated with WMC, LCOM was fairly correlated with WMC, NPM was correlated with WMC, and DIT was highly correlated with MFA. RFC was correlated with LOC, LOC was fairly correlated with AMC, and IC was strongly correlated with CBM. These correlations did not indicate that some metrics could be easily substituted by others. However, they were a good starting point to reduce the number of metrics used in the study.

Based on common knowledge about object-oriented metrics and the correlations studied, the following metrics were considered candidates to be overlooked or substituted by other metrics:
- WMC was correlated with RFC, LCOM, and NPM. The information conveyed by this metric was found also in LOC (the more methods in a

class, the more lines of codes) and RFC (which includes WMC in its computation).
- DIT was strongly correlated with MFA. This correlation was strong because DIT and MFA are measures of inheritance.
- RFC strongly correlated with WMC and LOC.
- LCOM was correlated with WMC. This correlation was strong because these measures are used to explore the cohesion of methods and attributes inside a class.
- IC was strongly correlated with CBM. This correlation was strong because a class is coupled to its parent class (in the case of IC) if one of its inherited methods is functionally dependent on the new or redefined methods, while CBM is the total number of new or redefined methods.
- We then analyzed the behavior of the following ten metrics: NOC, CBO, RFC, LCOM, Ca, Ce, LCOM3, MOA, MFA, CAM, IC, and CC.

### 5.2 Correlating Metrics with Bugs

To test the relationship between the metrics and the number of bugs, we conducted a correlation analysis. The correlation analysis is used to find the degree to which changes in the value of an attribute (one of the modularity measures) are associated with the changes in another attribute (the number of faults in a version).

If the measure tends to increase when the number of bugs increases, the Kendall correlation coefficient is positive. If the measure tends to decrease when the number of faults increases, the Kendall correlation coefficient is negative. Table 4 shows that CBO, RFC, Ce, MFA, and IC had low positive correlations with the number of bugs, whereas CAM had low negative correlations with the number of bugs.

### 5.3 Defect Density

In this subsection, we report the results of correlating the selected metrics with the number of bugs. Table 5 shows the defect densities found in the selected systems. Comparing the results obtained here and the numbers indicated in the literature [15, 16, 17], we can see that all selected open source systems have very low defect density which indicates a good quality products. jEdit has the best defect density rate (0.06) and comes second is Tomcat with (0.38). These two projects are very popular and widely used in several communities.

## 6. Related Work

Several previous researchers reported their answers to the question, "What is the typical defect density of a project?" Akiyama [15] reported that for each thousand lines of code (KLOC), there were 23 defects. McConnell [16] reported 1 to 25 defects, and Chulani [17] reported 12 defects.

The review of the relevant literature revealed several definitions of defect density. A recent overview study of defect density used the cumulative defects of all releases and the size of the last release to define defect density [13]. Their main argument was that the code base usually undergoes complex transformations, which makes it difficult to match a defect to the corresponding code base. In another study, Zhu and Faller [18] assessed defect density in evolutionary product development by using the aggregated churned LOC to measure size in calculating defect density. Their main argument was that the same code repository can have different numbers of defects regardless of whether those defects are in previous or future releases. Mohagheghi, et al. [19] studied a large, distributed system developed by Ericsson and compared the defect density of the system considering the re-used components and non-reused components. They found that reused components had lower defect density than the non-reused components. Raghunathan, et al. [20] compared the quality of open source, closed source software, and found no difference between them. Phipps [21] compared C++ and Java programs and found that C++ programs had two to three times as many defects per line of code as Java programs had.

In most of the related work, product metrics were used to study the proneness to defects without considering defect density. This gap in the literature indicates the need for research that characterizes product metrics based on defect density.

## 7. Conclusion

Building software that is of high quality is an essential aim for software engineering and information systems practitioners. To measure quality of software, different metrics are used and are available especially in open source software projects. Open source systems that are used in this study include Camel, Xalan, Tomcat, Ant, Xerces, jEdit, POI and Velocity. Many product metrics for the mentioned systems were used in this study including: weighted methods per class, depth of class, number of children, coupling between object classes, response for a class and others. This study has shown that defect density correlates disproportionally with open source software products and proportionally with the size of the product. Additionally, different metrics were found to be related to each other and bugs were found to be positively related to most metrics while only negatively related to cohesion among methods of class. Future work will focus on usage of more types of software metrics and building defect density prediction models.

Table 3. The Kendall rank cross-correlation coefficients of the considered metrics

|       | wmc   | dit   | Noc   | cbo   | rfc   | lcom  | ca    | ce    | npm   | lcom3 | loc   | dam   | moa   | mfa   | cam   | ic    | cbm   | amc   | avg   |
|-------|-------|-------|-------|-------|-------|-------|-------|-------|-------|-------|-------|-------|-------|-------|-------|-------|-------|-------|-------|
| wmc   | 1.00  | -0.02 | 0.17  | 0.34  | **0.69** | **0.65** | 0.24  | 0.26  | **0.80** | -0.26 | 0.52  | 0.35  | 0.37  | -0.16 | -0.65 | 0.18  | 0.19  | 0.17  | 0.41  |
| dit   | -0.02 | 1.00  | 0.00  | 0.08  | 0.08  | 0.00  | -0.14 | 0.25  | 0.00  | -0.03 | 0.11  | 0.04  | 0.02  | **0.82** | 0.01  | 0.58  | 0.53  | 0.15  | -0.11 |
| noc   | 0.17  | 0.00  | 1.00  | 0.18  | 0.13  | 0.15  | 0.26  | 0.09  | 0.13  | -0.06 | 0.09  | 0.11  | 0.14  | -0.02 | -0.12 | 0.02  | 0.02  | -0.01 | 0.07  |
| cbo   | 0.34  | 0.08  | 0.18  | 1.00  | 0.42  | 0.25  | 0.53  | 0.58  | 0.29  | -0.16 | 0.29  | 0.22  | 0.35  | 0.02  | -0.35 | 0.25  | 0.24  | 0.19  | 0.26  |
| rfc   | **0.69** | 0.08  | 0.13  | 0.42  | 1.00  | 0.46  | 0.17  | 0.40  | 0.53  | -0.29 | **0.71** | 0.37  | 0.40  | -0.04 | -0.59 | 0.26  | 0.26  | 0.43  | 0.45  |
| lcom  | **0.65** | 0.00  | 0.15  | 0.25  | 0.46  | 1.00  | 0.18  | 0.21  | 0.54  | 0.03  | 0.33  | 0.10  | 0.16  | -0.09 | -0.46 | 0.10  | 0.11  | 0.07  | 0.26  |
| ca    | 0.24  | -0.14 | 0.26  | 0.53  | 0.17  | 0.18  | 1.00  | 0.13  | 0.19  | -0.08 | 0.09  | 0.10  | 0.18  | -0.17 | -0.22 | -0.01 | -0.01 | -0.04 | 0.18  |
| ce    | 0.26  | 0.25  | 0.09  | 0.58  | 0.40  | 0.21  | 0.13  | 1.00  | 0.19  | -0.15 | 0.28  | 0.19  | 0.30  | 0.19  | -0.30 | 0.37  | 0.36  | 0.23  | 0.20  |
| npm   | **0.80** | 0.00  | 0.13  | 0.29  | 0.53  | 0.54  | 0.19  | 0.19  | 1.00  | -0.23 | 0.38  | 0.29  | 0.32  | -0.12 | -0.54 | 0.17  | 0.18  | 0.07  | 0.32  |
| lcom3 | -0.26 | -0.03 | -0.06 | -0.16 | -0.29 | 0.03  | -0.08 | -0.15 | -0.23 | 1.00  | -0.24 | -0.67 | -0.32 | 0.03  | 0.22  | -0.15 | -0.14 | -0.21 | -0.19 |
| loc   | 0.52  | 0.11  | 0.09  | 0.29  | **0.71** | 0.33  | 0.09  | 0.28  | 0.38  | -0.24 | 1.00  | 0.29  | 0.35  | 0.02  | -0.46 | 0.17  | 0.17  | **0.65** | 0.39  |
| dam   | 0.35  | 0.04  | 0.11  | 0.22  | 0.37  | 0.10  | 0.10  | 0.19  | 0.29  | -0.67 | 0.29  | 1.00  | 0.40  | -0.04 | -0.29 | 0.15  | 0.14  | 0.19  | 0.21  |
| moa   | 0.37  | 0.02  | 0.14  | 0.35  | 0.40  | 0.16  | 0.18  | 0.30  | 0.32  | -0.32 | 0.35  | 0.40  | 1.00  | -0.05 | -0.36 | 0.15  | 0.15  | 0.22  | 0.26  |
| mfa   | -0.16 | **0.82** | -0.02 | 0.02  | -0.04 | -0.09 | -0.17 | 0.19  | -0.12 | 0.03  | 0.02  | -0.04 | -0.05 | 1.00  | 0.12  | 0.48  | 0.45  | 0.14  | -0.20 |
| cam   | -0.65 | 0.01  | -0.12 | -0.35 | -0.59 | -0.46 | -0.22 | -0.30 | -0.54 | 0.22  | -0.46 | -0.29 | -0.36 | 0.12  | 1.00  | -0.18 | -0.19 | -0.19 | -0.32 |
| ic    | 0.18  | 0.58  | 0.02  | 0.25  | 0.26  | 0.10  | -0.01 | 0.37  | 0.17  | -0.15 | 0.17  | 0.15  | 0.15  | 0.48  | -0.18 | 1.00  | **0.92** | 0.16  | 0.10  |
| cbm   | 0.19  | 0.53  | 0.02  | 0.24  | 0.26  | 0.11  | -0.01 | 0.36  | 0.18  | -0.14 | 0.17  | 0.14  | 0.15  | 0.45  | -0.19 | **0.92** | 1.00  | 0.14  | 0.10  |
| amc   | 0.17  | 0.15  | -0.01 | 0.19  | 0.43  | 0.07  | -0.04 | 0.23  | 0.07  | -0.21 | **0.65** | 0.19  | 0.22  | 0.14  | -0.19 | 0.16  | 0.14  | 1.00  | 0.31  |
| avg   | 0.41  | -0.11 | 0.07  | 0.26  | 0.45  | 0.26  | 0.18  | 0.20  | 0.32  | -0.19 | 0.39  | 0.21  | 0.26  | -0.20 | -0.32 | 0.10  | 0.10  | 0.31  | 1.00  |

Table 4. Correlation coefficient of Metrics and Bugs

|      | NOC  | CBO  | RFC  | LCOM | Ca   | Ce   | LCOM3 | MOA  | MFA  | CAM   | IC   | CC   |
|------|------|------|------|------|------|------|-------|------|------|-------|------|------|
| Bugs | 0.06 | 0.12 | 0.13 | 0.09 | 0.04 | 0.21 | 0.03  | 0.09 | 0.11 | -0.14 | 0.12 | 0.09 |

Table 5. Defect Density Results

| *System* | *Defect Density* |
|----------|------------------|
| Camel    | 4.43             |
| Xalan    | 2.82             |
| Tomcat   | 0.38             |
| Ant      | 1.62             |
| Xerces   | 11.30            |
| jEdit    | 0.06             |
| POI      | 3.87             |
| Velocity | 3.33             |

**Mamdouh Alenezi** is the chairman of the computer science department in the Collage of Computer and Information Sciences at Prince Sultan University. He received his Ph.D. degree in Software Engineering from Department of Computer Science at North Dakota State University, Fargo, ND in 2014. He got a Master's degree from DePaul University and got a Bachelor's degree from Prince Sultan University. Dr. Alenezi taught many software engineering courses including Software Construction, Software Requirements, and Group Dynamics. His research interests include Mining Software Repositories, Software Maintenance, Software Testing, and Machine Learning. He has numerous publications in the field of Software Engineering and continually conducts reviews for many conferences in the same field.

**Ibrahim Abunadi** is Assistant Professor and website content director in the Collage of Computer and Information Sciences at Prince Sultan University in Saudi Arabia. He received his Ph.D. in Information Systems from the School Information Communication Technology at Griffith University in Australia. Dr. Abunadi taught many courses including Human Computer Interaction, Business Process Management, Enterprise Architecture, Technology Innovations, Business Analysis, Computer Databases and Computer Applications for Business. He has worked as an IT analyst for Computer Associates and as a strategic consultant for the Saudi Computer Association. His research focuses on software engineering, technology adoption, e-government and human-computer interaction. He has numerous publications in the field of Information Systems and Software Engineering and continually conducts reviews for many conferences and journals in the same fields. Dr. Abunadi is a member of the following associations: Saudi and Australian Computer Societies, ACM, Association of Information Systems and IEEE.